\documentclass{ptapap}

\author{Coralie Neiner}[LESIA]
\author{Gregg A. Wade}[RMC]
\author{Stephen C. Marsden}[USQ]
\author{Aurore Blaz\`ere}[ULG,LESIA]
\affil[LESIA]{LESIA, Observatoire de Paris, PSL Research University, CNRS,
Sorbonne Universit\'es, UPMC Univ. Paris 6, Univ. Paris  Diderot, Sorbonne
Paris Cit\'e, 5 place Jules Janssen, 92195 Meudon, France}
\affil[RMC]{Department of Physics, Royal Military College of Canada, PO Box
17000 Station Forces, Kingston, ON, Canada K7K 0C6}
\affil[USQ]{Computational Engineering and Science Research Centre, University of
Southern Queensland, Toowoomba 4350, Australia}
\affil[ULG]{Institut d'Astrophysique et de G\'eophysique, Universit\'e de
Li\`ege, all\'ee du 6 ao\^ut, 17, B\^at. B5c, 4000 Li\`ege, Belgium}

\title{The BRITE spectropolarimetric program}

\begin{document}

\maketitle

\begin{abstract}
A high-resolution spectropolarimetric survey of all (573) stars brighter than
magnitude V=4 has been undertaken with Narval at TBL, ESPaDOnS at CFHT, and
HarpsPol at ESO, as a ground-based support to the BRITE constellation of
nano-satellites in the framework of the Ground-Based Observation Team (GBOT).
The goal is to detect magnetic fields in BRITE targets, as well as to provide
one very high-quality, high-resolution spectrum for each star. The survey is
nearly completed and already led to the discovery of 42 new magnetic stars and
the confirmation of several other magnetic detections, including field
discoveries in, e.g., an Am star, two $\delta$\,Scuti stars, hot evolved stars,
and stars in clusters. Follow-up spectropolarimetric observations of
approximately a dozen of these magnetic stars have already been performed to
characterise their magnetic field configuration and strength in detail.
\end{abstract}

\section{Introduction}

Magnetic fields have a significant impact on internal stellar structure through
their role in rotation, the transport of angular momentum, mixing, and mass
loss. They also impact the circumstellar environment, e.g. through confinement
of the wind particles into a magnetosphere. As a consequence, knowing the
magnetic properties of the BRITE targets is crucial to interpret the BRITE data
correctly.

In particular, a magnetic field produces a splitting of pulsation modes. The
observed magnetically split multiplets depend on the field strength and on the
obliquity of the field compared to the rotation axis \citep[see,
e.g.,][]{shibahashi2000}. Therefore, knowledge of the magnetic properties is
fundamental for proper pulsation mode identification for seismic modelling. Not
knowing that the star is magnetic could lead to misinterpreting pulsation
multiplets as high-order modes instead of low-order magnetically-split modes
and, thus, to an incorrect derivation of the internal stellar structure.

In addition, the presence of a magnetic field inhibits mixing inside the star.
This was observed, for example, in the $\beta$\,Cep star V2052 Oph. In this star
seismic models were found to require no overshooting in spite of the relatively
high rotational velocity \citep{briquet2012}. This could be explained by the
presence of a $\sim$400 G dipolar field \citep{neiner2012}, since this field
strength is well above the critical field limit at which theory predicts a
magnetic inhibition of mixing in this star \citep{zahn2011}. This example shows
how knowing magnetic properties can yield improved constraints for seismic
modelling. 

Magnetic fields can also be present in other types of pulsators. For example,
the recent discovery of magnetic $\delta$\,Scuti Ap stars
\citep{neinerlampens2015, escorza2016}, or the class of rapidly oscillating Ap
(roAp) stars, require the consideration of magnetic effects in the study of the
pulsations in these types of stars.

As a consequence, we endeavoured to perform a complete spectropolarimetric
survey of all the 573 brightest (V$\leq$4) targets in the sky and characterise
those for which a magnetic field is detected. This survey will also provided a
complete magnitude-limited sample for the statistical analysis of the occurence
of magnetic fields in stars.

\section{The BRITE spectropolarimetric survey}

The BRITE spectropolarimetric survey has been performed on the three
high-resolution stellar spectropolarimeters available in the world: Narval at
TBL (T\'elescope Bernard Lyot) in France, ESPaDOnS at CFHT (Canada-France-Hawaii
Telescope) in Hawaii, and HarpsPol at La Silla in Chile. Narval was used for all
stars with a declination above -20$^\circ$, ESPaDOnS for those between
-45$^\circ$ and -20$^\circ$, and HarpsPol for the targets below -45$^\circ$. 
High-quality spectropolarimetric data were already available in archives for 87
stars among the 573 targets. Therefore, we needed to observe only the remaining
486 stars.

\begin{figure}
\includegraphics[width=0.65\textwidth, angle=-90, clip]{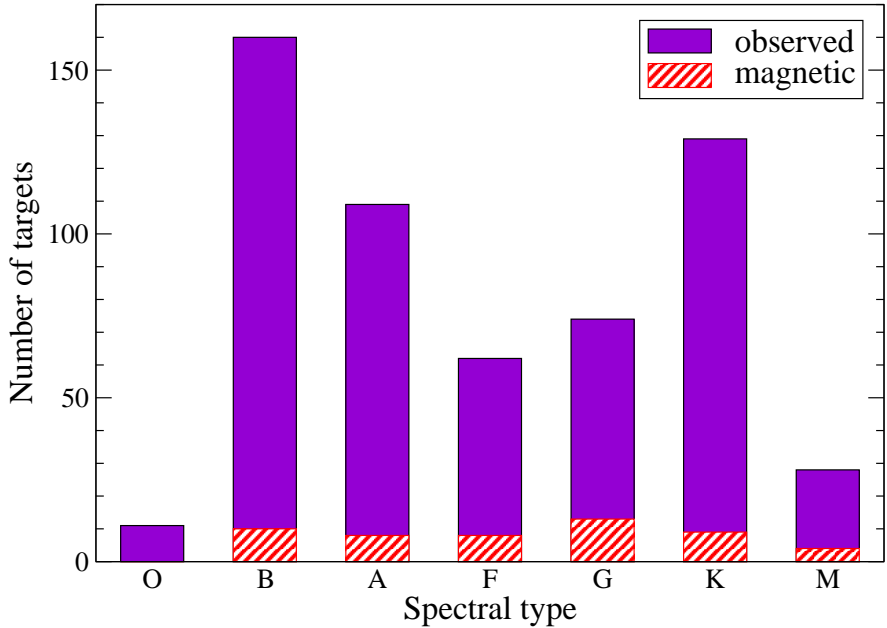}
\caption{Number of stars in the sample (full purple bars) and number of detected
magnetic stars (hatched red bars) per stellar type.}
\label{BritePoltargets}
\end{figure}

About 50\% of the targets in the sample are hot stars, while the other 50\% are
mostly evolved cool stars (see Fig.~\ref{BritePoltargets}). Hot
stars host fossil magnetic fields, which are usually simple (mostly dipolar) in
structure, stable, and rather strong (above 300 G at the poles), but appear in
only about $\sim$10\% of hot stars \citep{grunhut2015, neiner2015}. On the
contrary, cool stars host dynamo fields similar to our Sun, which are produced
contemporaneously by the star itself, are variable on short timescales, very
weak on average (but can be stronger in spots), and observable in most cool
stars. Therefore, we separated the sample in two groups: hot stars between
spectral types O and F5, for which we aimed at detecting longitudinal field
strengths with error bars of the order of 5 G (i.e. dipolar field strength above
50 G), and cool stars with spectral types between F5 and M, for which the
desired precision was 10 times better. 

As of mid-August 2016, we have obtained 129 hours of Narval queue service time,
56 hours of ESPaDOnS queue service time, and 23.2 nights (equivalent to 210
hours) in visitor mode of HarpsPol time. In addition, a large amount of
observations were executed using C-time with Narval and ESPaDOnS, as the BRITE
targets are excellent queue fillers at telescopes. In total we observed 464
stars, which corresponds to a $\sim$95\% completion of the survey (as of
mid-August). 

Data were collected in circular polarisation (Stokes V) mode and the
signal-to-noise ratio in the intensity spectrum is above 1000 for all targets,
and often above 2000. We obtained one measurement per star and applied the Least
Square Deconvolution (LSD) technique \citep{donati1997} with template masks
extracted from VALD \citep{piskunov1995, kupka1999} to search for magnetic
signatures. If a Stokes V signature was detected, we usually obtained a second
measurement to confirm the magnetic detection.

52 stars were found to be magnetic, which corresponds to a 11\% detection rate,
including 42 stars that were not known to host a magnetic field before this
survey (see Fig.~\ref{BritePoltargets}). This includes 4 B and A supergiant
stars, 14 other B and A stars, 7 F, G, and K dwarfs or subwarfs, 23 F, G, and K
giants or supergiants, and 4 M giants. While the detection rate of 9\% in hot
stars corresponds to the known $\sim$10\% occurence rate of magnetism in these
objects \citep{grunhut2015}, the rate of detections in cool stars (11\%) is
lower than expected for these stars \citep[see][]{petit2014}. This rate
increases to 19\% when considering only class V stars of F, G, and K types.
Therefore, the low detection rate is probably partly due to the fact that most
cool stars in the sample are evolved and thus their field is weaker than on the
main sequence, and partly due to the fact that fields are variable and can be
below our detection threshold at the time of observation.

The most interesting magnetic targets were then transferred to the follow-up
part of the program for a full magnetic characterisation.

\section{Spectropolarimetric follow-up of magnetic BRITE targets}

When a star is found to be magnetic in the survey, multiple spectropolarimetric
observations are acquired at various phases of the rotation cycle. In this way,
it is possible to observe the magnetic field under various angles and, using the
rotational modulation of the Stokes V profile and longitudinal field value, it
is possible to reconstruct the magnetic field configuration on the stellar
surface (such as the obliquity angle between the rotation axis and magnetic
axis), its strength, and obtain additional stellar parameters (e.g., the
rotation period and the inclination angle of the star's rotation axis with
respect to the observer).

As of mid-August, such a spectropolarimetric follow-up had already been
performed for 11 magnetic BRITE targets, and was ongoing for 7 others. Among
these 18 targets, 13 are B and A stars, and 5 are F and G stars. Four of the 18
targets are binary systems. Follow-up observations will continue in 2016 and
2017. 

\subsection{The B3V star i\,Car}

\begin{figure}
\includegraphics[width=0.65\textwidth, angle=-90, clip]{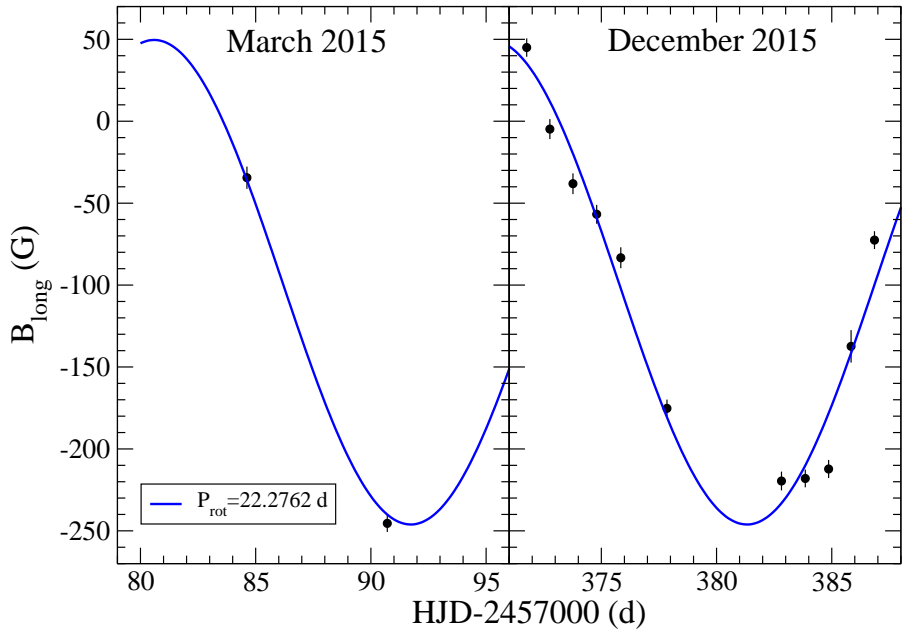}
\caption{Longitudinal field measurements (black dots) of the B3V star i\,Car
obtained with HarpsPol, with a dipole fit (blue curve).}
\label{icar}
\end{figure}

Fig.~\ref{icar} shows an example of a hot star, the B3V star i\,Car, detected to
be magnetic in the BRITE spectropolarimetric survey with two measurements
gathered in March 2015 \citep{neiner2015_letter}. In December 2015 we acquired
11 additional measurements and could determine its dipole field strength
($\sim$1 kG) and rotation period (P$_{\rm rot}\sim$22.28 d). The rotation period
can be precisely determined thanks to the two datapoints obtained in March 2015.

A dipole fit is shown in Fig.~\ref{icar}. Deviation between this fit and the
data could indicate that the field includes a quadrupole component, or that the
shape of the longitudinal field curve is modified by surface abundance patches.

\subsection{The Am star Alhena}

Another interesting example of a hot star detected to be magnetic in the BRITE
spectropolarimetric survey is the Am star Alhena \citep{blazere_alhena}. While
it was recently discovered that all Am stars studied with sufficient precision
host ultra-weak magnetic fields with a peculiar magnetic signature shape
\citep{blazere_Am}, Alhena is the first example of a magnetic Am star showing a
standard Zeeman signature.

\begin{figure}
\includegraphics[width=0.65\textwidth, angle=-90, clip]{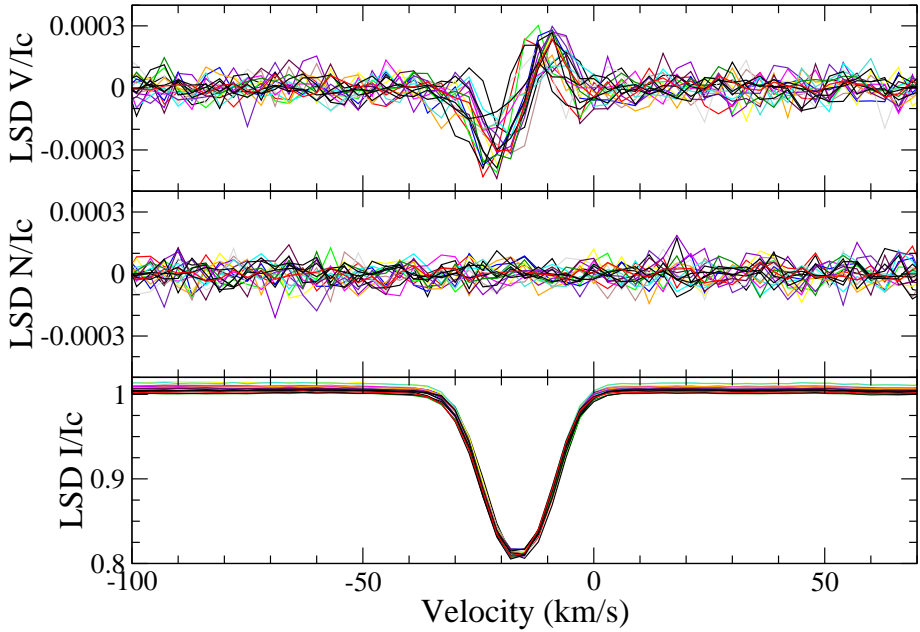}
\caption{LSD Stokes I (bottom), null polarization (middle), and Stokes V (top)
profiles of the Am star Alhena, obtained with Narval. The Stokes V profiles show
standard Zeeman signatures, contrary to other Am stars.}
\label{alhena}
\end{figure}

Fig.~\ref{alhena} shows 20 LSD profiles of Alhena taken with Narval between
October 2014 and April 2016. The variation of the Stokes V profile is small over
the 1.5 years of observations, suggesting that Alhena is seen under a specific
geometrical configuration. 

Moreover, its field, although weak (longitudinal field of a few gauss), is
stronger than the field of the other magnetic Am stars (less than 1 G). The
difference could be related to the lower level of microturbulence in Alhena 
\citep{blazere_alhena}. Indeed, the peculiar Stokes V shape observed in the
other Am stars is thought to be due to supersonic convection flows in the
shallow convective shell, which could be the source of sharp velocity and
magnetic gradients \citep{blazere_Am}. 

\subsection{The F7V star $\chi$\,Dra}

\begin{figure}
\includegraphics[width=0.65\textwidth, angle=-90, clip]{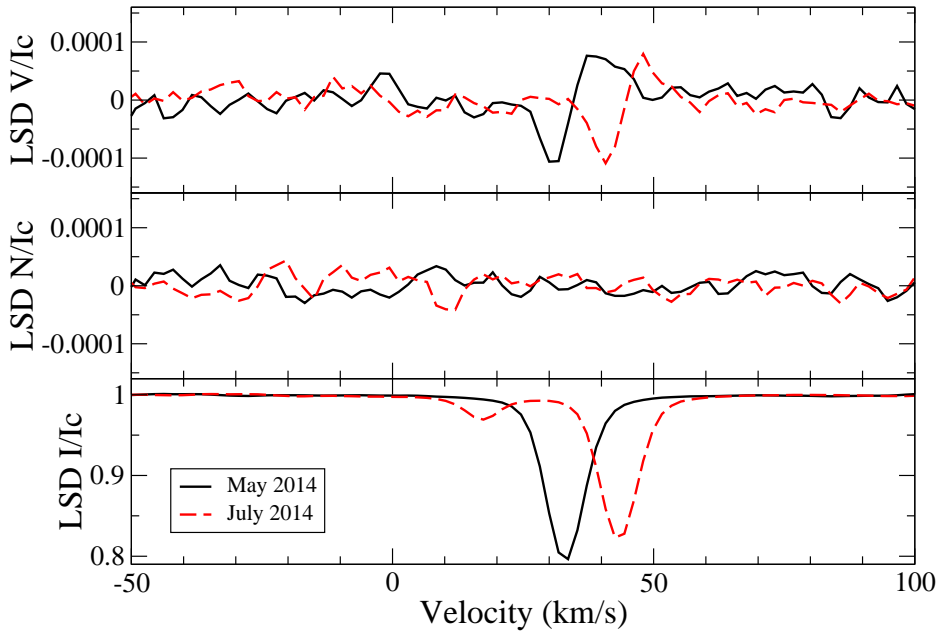}
\caption{LSD Stokes I (bottom), null polarization (middle), and Stokes V (top)
profiles of the F7V+K system $\chi$\,Dra, obtained with Narval. In the
observation collected in May (in solid black) the lines of both stars are
superimposed, while the observation gathered in July (in dashed red) clearly
shows that the F7V star is the magnetic one.}
\label{chidra}
\end{figure}

Fig.~\ref{chidra} shows an example of a cool star: the magnetic detection in the
binary system $\chi$\,Dra obtained during the survey in May and July 2014. In
the first observation the spectral lines of the two stars are superimposed and
thus the LSD profiles do not allow to distinguish the two components.
Fortunately, the radial velocity of the two stars were different in the second
observation and the LSD profiles clearly show that the F7V star is magnetic,
while its cooler companion is not (with our detection threshold). Longitudinal
field values measured for $\chi$\,Dra are of the order of a few gauss.

Preliminary modelling of series of follow-up data obtained in 2015 and 2016
indicates that the rotation period is of the order of 5 days and show that the
field of $\chi$\,Dra is relatively simple with a strong poloidal component. Such
simple configurations are not common in cool stars. $\chi$\,Dra might be located
near the transition between fossil and dynamo fields. 

\subsection{The M giants}

\begin{figure}
\includegraphics[width=0.35\textwidth, angle=-90, clip]{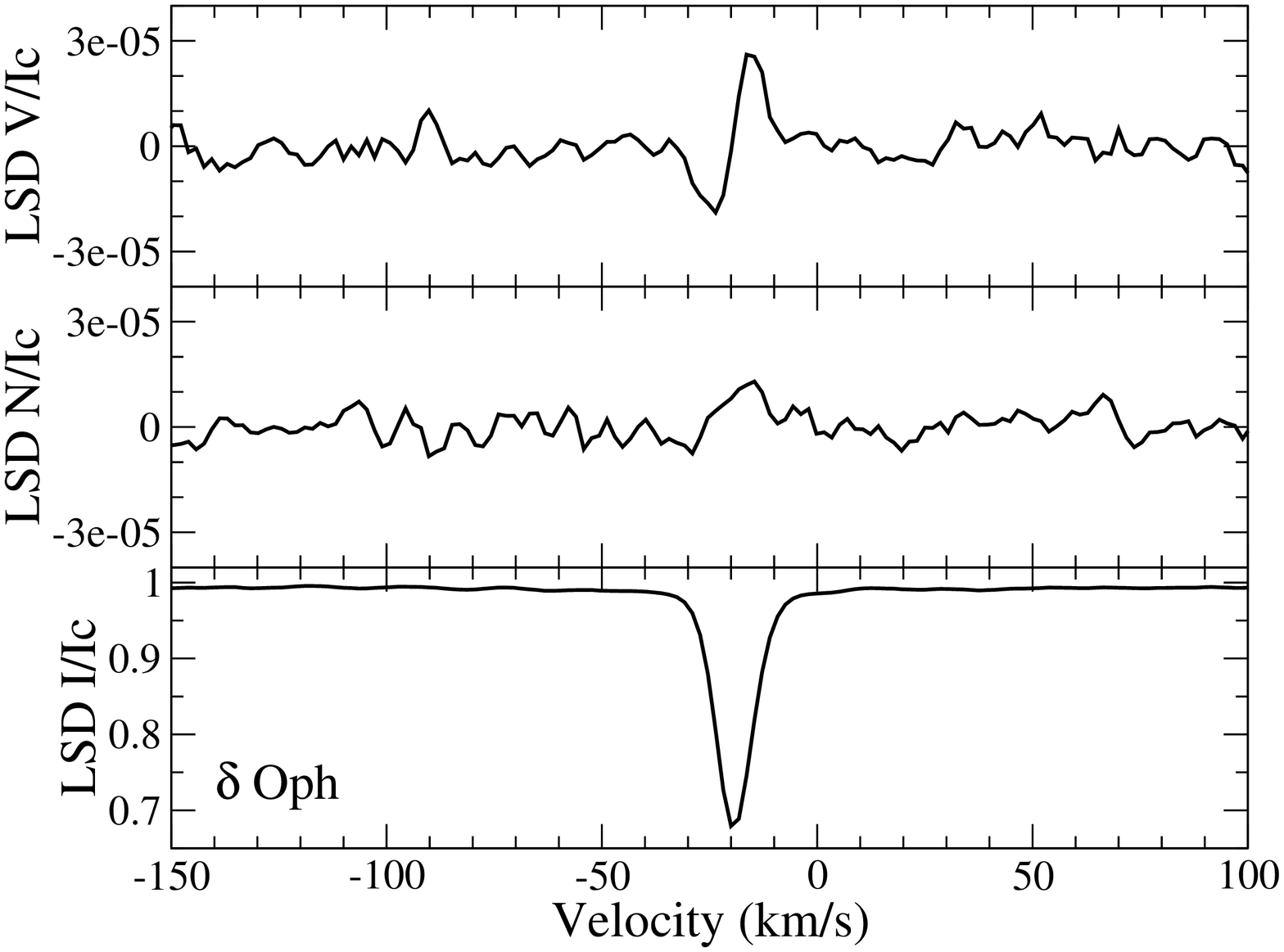}
\includegraphics[width=0.35\textwidth, angle=-90, clip]{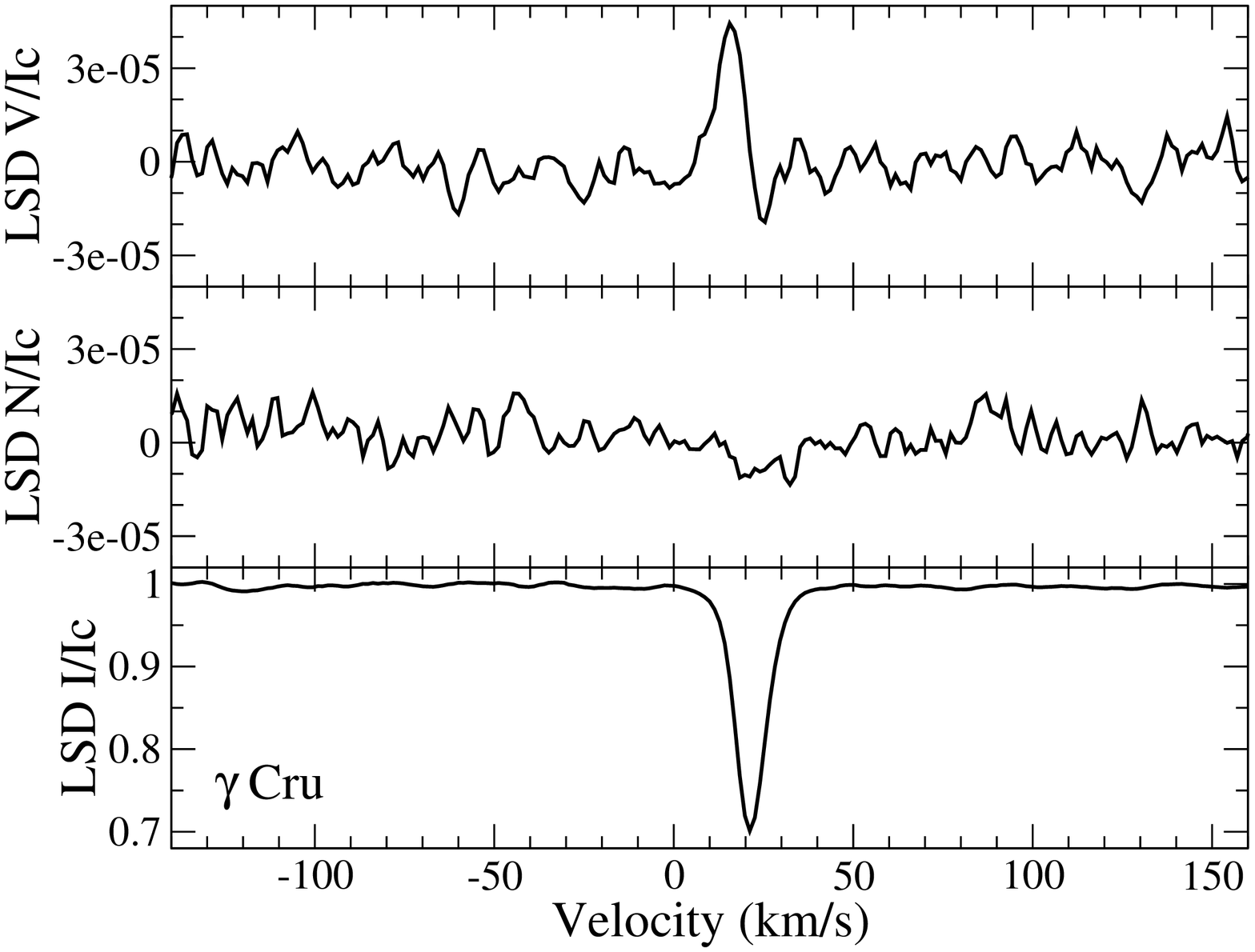}\\
\includegraphics[width=0.35\textwidth, angle=-90, clip]{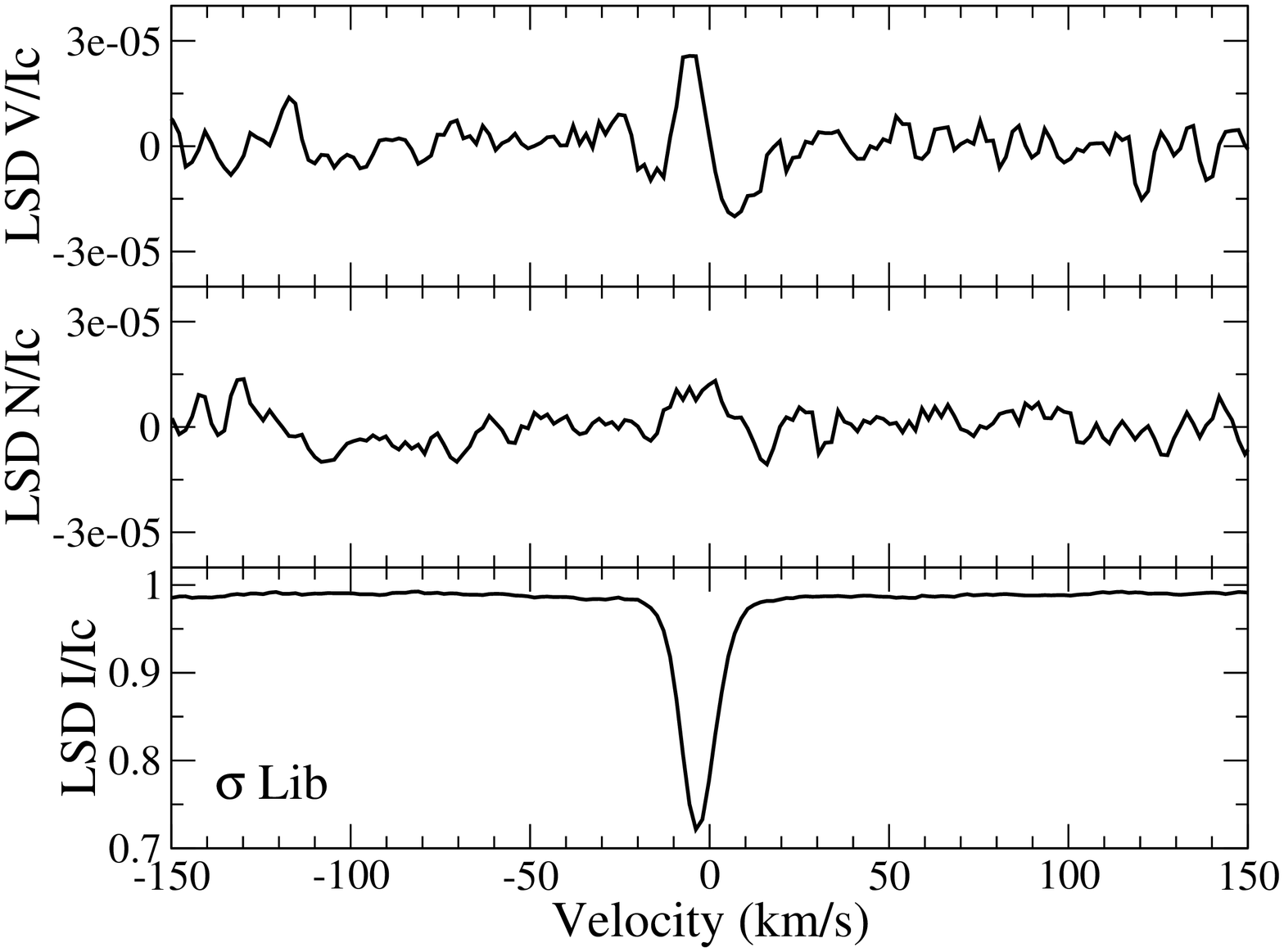}
\includegraphics[width=0.35\textwidth, angle=-90, clip]{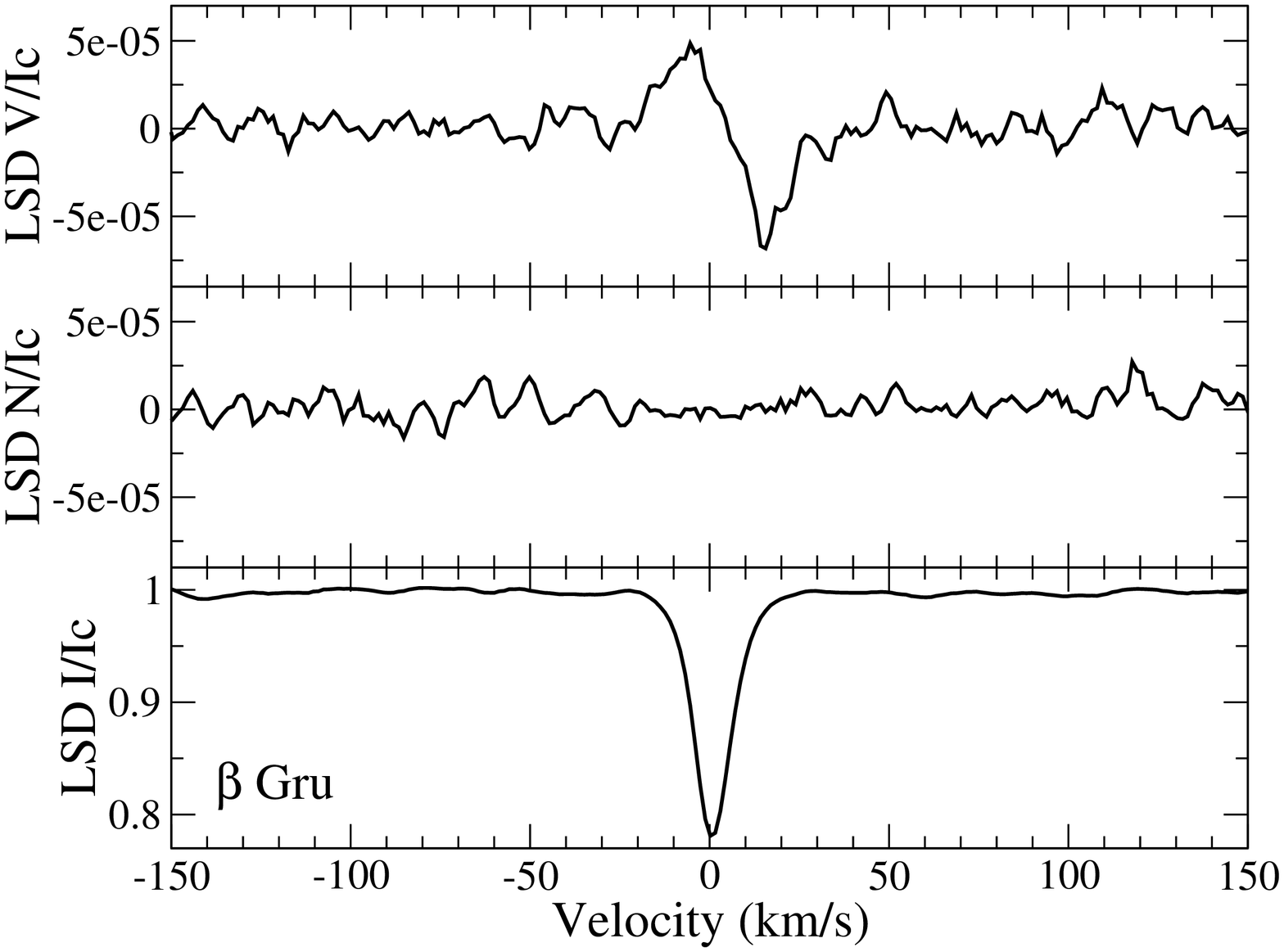}
\caption{LSD Stokes I (bottom), null polarization (middle), and Stokes V (top)
profiles of the four M giants detected to be magnetic in the survey.}
\label{Mgiants}
\end{figure}

A magnetic detection was obtained in four M giants in our sample: $\delta$\,Oph
(M0.5III), $\gamma$\,Cru (M3.5III), $\sigma$\,Lib (M3/4III), and $\beta$\,Gru
(M5III). $\beta$\,Gru and $\gamma$\,Cru were observed with HarpsPol,
$\sigma$\,Lib with ESPaDOnS, and $\delta$\,Oph with Narval. A magnetic signature
is clearly visible in these four targets, as shown in Fig.~\ref{Mgiants}.

\section{Conclusion}

We have almost completed the spectropolarimetric survey of all 573 stars
brighter than V=4, using all three high-resolution stellar spectropolarimeters
available in the world. We acquired one high-resolution, high signal-to-noise
spectrum of each target, for which no adequate data was available in the
archives yet. We detected 52 magnetic stars, including 42 new ones.

Follow-up observations of the most interesting magnetic stars allow us to
characterise their magnetic field in detail. This will allow us to provide
crucial information for the interpretation of the BRITE photometric data, as
well as strong constraints for seismic modelling of the pulsating stars. This
combined technique, called magneto-asteroseismology, is developing fast
\citep{neinerIAU,mathis2015} and is particularly appropriate for bright stars
which can be easily studied with spectropolarimetry even when they host weak
magnetic fields. 

Finally, for all stars including those that are not magnetic, the exquisite
spectra collected in this program can be used to determine stellar parameters
and chemical abundances.

\acknowledgements{CN thanks the French program for stellar physics (PNPS) for
its financial support to the BRITE spectropolarimetric program, as well as the
teams at TBL and CFHT for their efficient and flexible observation scheduling.}

\bibliographystyle{ptapap}
\bibliography{Innsbruck_Neiner}

\end{document}